\documentclass[%
 aip,
 amsmath,amssymb,
preprint,%
]{revtex4-1}

\draft 

\usepackage{graphicx}
\usepackage{dcolumn}
\usepackage{bm}

\usepackage[utf8]{inputenc}
\usepackage[T1]{fontenc}
\usepackage{mathptmx}

\begin{document}
\preprint{}

\title{Interface studies of molecular beam epitaxy (MBE) grown ZnSe-GaAs heterovalent structures}

\author{Zongjian Fan}
\email[]{zjfan@ucdavis.edu}
\affiliation{ 
Department of Electrical and Computer Engineering, University of California, Davis, Davis, CA 95616
}

\author{Krishna Yaddanapudi}
\email[]{grkyaddanapudi@ucdavis.edu}
\affiliation{ 
Department of Materials Science and Engineering, University of California, Davis, Davis, CA 95616
}

\author{Ryan Bunk}
\affiliation{ 
Department of Electrical and Computer Engineering, University of California, Davis, Davis, CA 95616
}

\author{Subhash Mahajan}
\affiliation{
Department of Materials Science and Engineering, University of California, Davis, Davis, CA 95616
}

\author{Jerry M. Woodall}
\affiliation{ 
Department of Electrical and Computer Engineering, University of California, Davis, Davis, CA 95616
}

\date{\today}
\begin{abstract}
Comprehensive investigations on ZnSe/GaAs and GaAs/ZnSe interfaces were carried out by photoluminescence (PL) and transmission electron microscopy (TEM), as a part of realizing high quality ZnSe-GaAs (100) hetero-valent structures (HS). The nature of ZnSe/GaAs interface under different surface terminations of GaAs was examined. The ZnSe/Ga-terminated GaAs was found to have a superior optical and microstructural quality, with a chemical interface consisting of a mixture of both the GaAs and ZnSe atomic constituents. For GaAs/ZnSe interface studies, a low-temperature migration enhanced epitaxy (LT-MEE) growth technique was used to grow GaAs layers under the conditions compatible to the growth of ZnSe. Both Ga and As-initialized LT-MEE GaAs/ZnSe interfaces were investigated. A defective transition layer was observed along the As-initialized GaAs/ZnSe interface, which may be attributed to the formation Zn$_3$As$_2$ compound. The correlation between the observed optical as well as structural properties of both the (GaAs/ZnSe and ZnSe/GaAs) interfaces and the growth conditions used in this study are discussed in detail. This study could provide a valuable insight on the interface nature of ZnSe-GaAs HS.\\
\end{abstract}
\maketitle
\section{\label{sec:level1}Introduction}
Zinc selenide (ZnSe) - gallium arsenide (GaAs) hetero-valent structure (HS) has long been recognized as a potential candidate for fabricating various optoelectronic devices including displays, light emitting diodes, and lasers with optimal chromaticity\cite{int1, int2, int3, int4, int5, int6}. Although the ZnSe-GaAs HS exhibits a low lattice mismatch (0.25\% at room temperature), the chemical valence mismatch at the ZnSe-GaAs HS interface is a bottleneck to practically integrate these two materials with tolerable interface defect density\cite{int4, int5, int6}. The chemical valance mismatch introduces defect states in the energy band gap of ZnSe-GaAs HS system, which could degrade the radiative emission efficiency of devices\cite{int1, int2, int3, int4, int7}. Moreover, these interface imperfections were found to be responsible for the high stacking fault density in ZnSe-GaAs HS, which could further degrade the lifetime of devices\cite{int8}.\\
In this regard, extensive investigations were carried out by various authors to understand the nature of ZnSe-GaAs HS interfaces, including the widely studied ZnSe on GaAs HS (ZnSe/GaAs interface)\cite{int9, int10, int11, int12, int13, int14, int15, int16, int17, int18, int19, int20} and the less investigated GaAs on ZnSe HS (GaAs/ZnSe interface)\cite{int4, int20}. The detailed investigations using capacitance-voltage (CV) and photocurrent spectra (PS) techniques revealed that the ZnSe/Ga-terminated GaAs interface contains the lowest interface defect state density compared to ZnSe/As-terminated GaAs\cite{int9, int10, int11, int14}.Though, the ZnSe/Ga-terminated GaAs interface is electrically good, structurally it was found to contain an ultra-thin Ga-Se (Ga$_2$Se$_3$) compound transition layer\cite{int17}, which could promote 3D nucleation of subsequently grown ZnSe, resulting in bulk ZnSe with high defect density\cite{int16}. In contrast, the electrically poor ZnSe/As-terminated GaAs was reported to contain a transition layer of Zn-As (Zn$_3$As$_2$)\cite{int17}, which could promote 2D nucleation of ZnSe resulting in bulk ZnSe with low defect density\cite{int16}. The formation of these compound Ga$_2$Se$_3$ and Zn$_3$As$_2$ ultra-thin transition layers may be attributed to the reaction between Ga (Zn) and Se (As) atoms during the initial stages of epitaxial deposition of ZnSe on Ga and As-terminated GaAs surfaces\cite{int13, int17, int16, int9}. There were a few reports on the stoichiometry of GaAs (100) surfaces intermediate between Ga-rich and As-rich extremes, which could promote 2D growth mode of subsequent ZnSe with improved optical and electrical properties\cite{int14, int20}.\\
In contrast to the widely studied ZnSe/GaAs interface, the GaAs/ZnSe interface was rarely investigated due to the fact that the underlying ZnSe decomposes at temperatures lower than the typical growth temperatures of GaAs using MBE (around $580^\circ$C)\cite{int2, int20}. In order to examine the GaAs/ZnSe interface, therefore, the GaAs layers have to be grown at typical growth temperatures of ZnSe (i.e. around $300^\circ$C), where the GaAs layers usually have poor crystalline quality. It was reported that the microstructure of GaAs layers grown on Zn-rich ZnSe at $300^\circ$C by MBE was observed to contain antiphase domains, which could degrade the surface quality of GaAs layers\cite{int20}. In this contribution, Migration Enhanced Epitaxy (MEE) technique was used to develop GaAs epitaxial layers at temperatures as low as $200^\circ$C with adequate crystalline quality\cite{int21, int22}. Nevertheless, the ZnSe/GaAs/ZnSe quantum structures developed using MEE were only able to deliver desirable photoluminescence (PL) at extremely low temperatures due to poor interface quality\cite{int3, int23, int24, int25, int26}. Moreover, the surface termination of ZnSe and the initial growth layer of GaAs were proven to affect the GaAs/ZnSe interface quality\cite{int4, int14, int15, int16, int17, int18}. Understanding the structural and optical properties of GaAs/ZnSe interface in relation to the widely studied ZnSe/GaAs interface is, therefore, essential to realize high quality ZnSe-GaAs HS systems.\\
We therefore revisit the growth of ZnSe-GaAs HS system with the aim of understanding the structural as well as the optical nature of ZnSe/GaAs and GaAs/ZnSe interfaces. The interfaces of ZnSe layers grown on MBE grown GaAs as well as MEE grown GaAs buffer layers were investigated in detail. Subsequently, the effect of the initial growth layer of GaAs on the quality of GaAs/ZnSe interfaces was examined. Room temperature (RT) PL and transmission electron microscopy (TEM) techniques were used to evaluate the optical as well as structural quality of these hetero-valent interfaces. An attempt has been made to establish a systematic correlation between the observed properties of these interfaces and the growth conditions used in this study.\\

\section{Experimental}
\begin{table*}
\caption{\label{tab:table1}Stack structures, growth methods, and growth conditions of the samples used in this study.}
\begin{ruledtabular}
\begin{tabular}{cccc}
Sample ID&Stack Structure&Growth Temperatures&V/III Ratio of GaAs\\
\hline
A&ZnSe/Ga-terminated GaAs&$300^\circ$C/$580^\circ$C&$\sim15$\\
B&ZnSe/As-terminated GaAs&$300^\circ$C/$580^\circ$C&$\sim15$\\
C&ZnSe/LT-MBE GaAs&$300^\circ$C/$300^\circ$C&$\sim1$\\
D&ZnSe/LT-MEE GaAs&$300^\circ$C/$300^\circ$C&\\
E&Ga-initialized GaAs/ZnSe&$300^\circ$C/$300^\circ$C&\\
F&As-initialized GaAs/ZnSe&$300^\circ$C/$300^\circ$C&\\
\end{tabular}
\end{ruledtabular}
\end{table*}
The ZnSe and GaAs layers in this study were grown in a single chamber Varian Gen II MBE system. The substrates used were n-type GaAs with (100) orientation and exhibit an etch pit density (EPD) < 500 cm$^{-2}$. Figure 1a and 1b shows the typical stack structures of ZnSe/GaAs and GaAs/ZnSe hetero-structures used in this study. Prior to growth, the substrates were baked in the load-lock chamber at a temperature of  $300^\circ$C for three hours to remove the water vapor. Subsequently, the GaAs substrates were treated thermally in the growth chamber at $\sim610^\circ$C under As overpressure to remove the native substrate oxide layer. The thermal treatment process was continued until a dot diffraction pattern was confirmed by Reflection High Energy Electron Diffraction (RHEED). In order to provide a better quality growth surface for ZnSe growth, a 500 nm thick Si-doped GaAs buffer layer was grown on oxide removed GaAs substrates at $580^\circ$C, and at a V/III flux ratio of $\sim15$. Steaky RHEED with $(2\times4)$ As-rich pattern was observed during GaAs buffer layer growth.\\
\begin{figure*}
\includegraphics[width=0.9\textwidth]{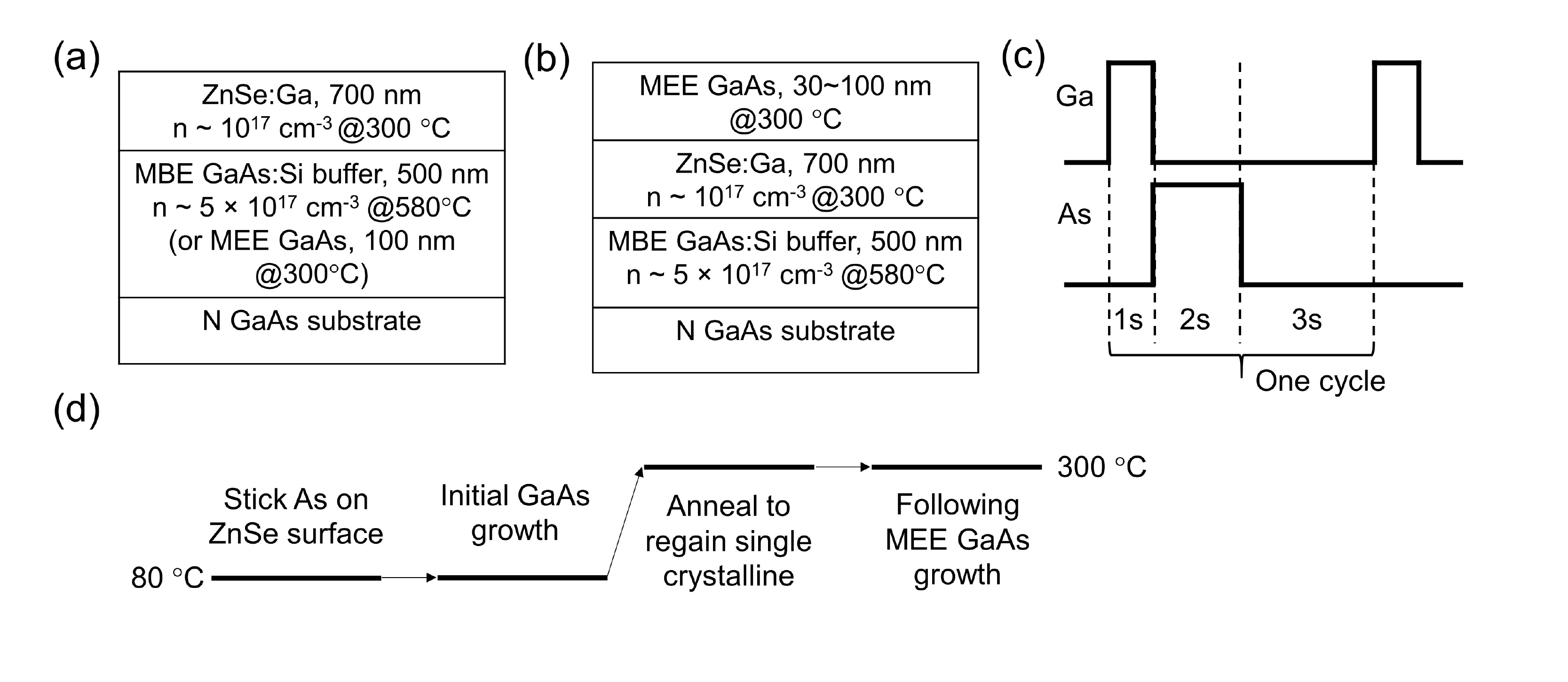}
\caption{\label{fig:epsart} Sample stack structures for (a) ZnSe/GaAs interface study, (b) GaAs/ZnSe interface study and (c) MEE growth process schematic diagram of each one cycle used in this work. (d) Illustration of procedure used to grow As-initialized GaAs on ZnSe surface.}
\end{figure*}
For ZnSe/GaAs interface study, $\sim700$ nm thick Ga-doped ZnSe layers were grown at $300^\circ$C on different surface terminated GaAs buffer layers in the same MBE growth chamber (Figure 1a). The ZnSe growth was initiated only after As background pressures brought by GaAs stabilized low enough in the growth chamber to prevent ZnSe from unwanted doping effects. The background pressure in the chamber was constantly monitored by residual gas analyzer (RGA). An effusive cell with compound ZnSe source material was used to provide the beam flux for ZnSe growth\cite{int27}.  A $c(2\times2)$ Zn-rich RHEED patter was observed during the  growth of ZnSe. Prior to ZnSe growth, the surface structures of GaAs buffer such as the as-grown $(2\times4)$ As-rich surface and $(4\times2)$ Ga-rich surface of GaAs buffers were confirmed by RHEED. The $(4\times2)$ Ga-rich surface was obtained via thermal annealing of GaAs buffers at $560^\circ$C, and in the absence of As environment. For comparison purpose, a ZnSe/GaAs hetero-structure was grown, where GaAs buffer was grown at low temperatures (LT) $300^\circ$C and at V/III $\sim1$. The lower V/III $\sim1$ was used to prevent excess As incorporation in bulk GaAs at LTs, as excess As is known to suppress the radiative recombination\cite{int28}. \\
In order to develop high quality GaAs buffer at LTs for the development of ZnSe-GaAs HS. We initially optimized the LT-MEE GaAs growth conditions and grown a ZnSe/LT-MEE GaAs hetero-structure for a comparative study. The LT-MEE GaAs layers were grown at $300^\circ$C with fixed Ga amount (1 monolayer (ML)) and varying As amount per cycle\cite{int21, int22} using both As$_4$ and As$_2$(see Figure 1c). The relationship between the number of surface sites and the beam flux pressure was derived from elsewhere\cite{int21, int22}. For GaAs/ZnSe interface study, $\sim100$ nm thick LT-MEE GaAs layers were grown on $c(2\times2)$ Zn-rich ZnSe at $300^\circ$C. Both the Ga-initialized and As-initialized LT-MEE GaAs layers were grown on ZnSe (see Figure 1b). The Ga-initialized GaAs growth was initiated by depositing a Ga ML on $c(2\times2)$ Zn-rich ZnSe surface at $300^\circ$C. In contrast, the As-initialized LT-MEE GaAs growth involves various steps: initiating the GaAs deposition at low temperatures below $80^\circ$C to achieve high As sticking coefficient on ZnSe surface\cite{int3}, annealing the grown GaAs at $300^\circ$C to obtain single crystal RHEED pattern, and then continuing further MEE GaAs growth at $300^\circ$C. The processing cycle used for As-initialized MEE GaAs growth is described in Figure 1d. The list of samples that have been used in this study were tabulated in Table 1 along with the growth conditions used.\\
The grown ZnSe-GaAs HS samples were characterized ex-situ using PL and TEM. All PL characterizations were measured by 405 nm excitation laser with maximum power of $\sim51$ mW at RT. For different incident laser power measurements, the laser power was altered by neutral density filters and calibrated by power meter. The microstructural and interface investigations were carried out using JEOL JEM 2100F-AC TEM, operating at an accelerating voltage 200 kV. The cross-sectional TEM lamella in this study were prepared using FEI Scios dual-beam focused ion beam (FIB).\\

\section{Results}
\subsection{\label{sec:level2}ZnSe/GaAs interface:}
\subsubsection{\label{sec:level3}Photoluminescence investigations:}
The investigations in this study were started from the relatively well-studied ZnSe/GaAs interface. Figure 2 shows the  band-edge PL emission spectra obtained from the ZnSe layers grown under identical conditions on different GaAs buffers, including the Ga-terminated GaAs (sample A), As-terminated GaAs (sample B), and LT-MBE GaAs (sample C) in Table I. The surface termination of GaAs buffers was confirmed by RHEED, prior to the growth of ZnSe, including the $(2\times4)$ As-rich and $(4\times2)$ Ga-rich GaAs surface. Since a large portion of GaAs PL signal will come from the GaAs substrate, which could bring ambiguity to this study, ZnSe PL was used to measure the interface quality.\\
\begin{figure*}
\includegraphics[width=0.9\textwidth]{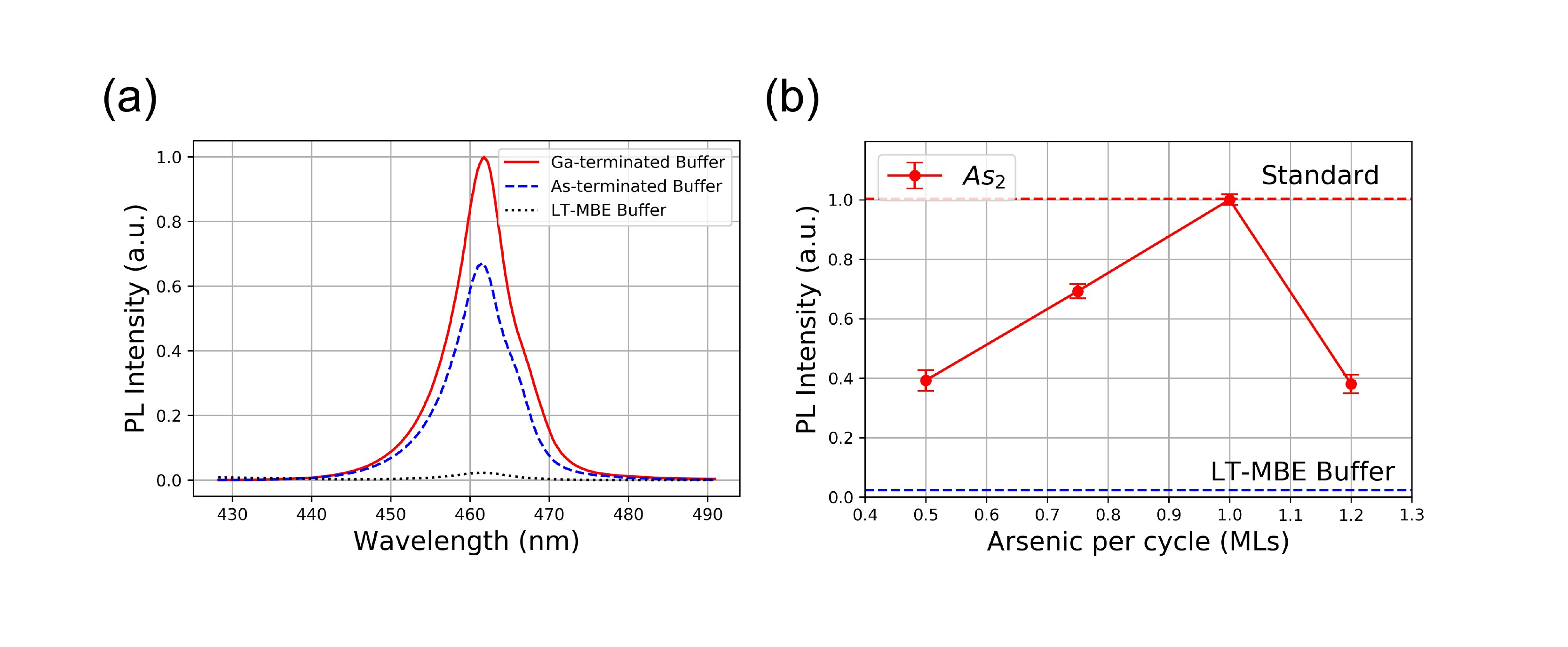}
\caption{\label{fig:epsart} (a) RT PL responses of ZnSe layers on different MBE GaAs buffer layers (sample A, B, C) and (b) RT PL peak responses intensity of ZnSe layers on different LT-MEE GaAs buffer layers. The LT-MEE GaAs layer were grown with different Arsenic amounts per cycle.}
\end{figure*}
As expected, the ZnSe layer grown on Ga-terminated GaAs (sample A) exhibits a high intense band-edge emission when compared to ZnSe/As-terminated GaAs (sample B) and ZnSe/LT-MBE GaAs (sample C) in Table I.  Although, the growth conditions of ZnSe layers were identical in this study, it appears that the surface termination of GaAs buffer strongly influences the quality of ZnSe layers and hence the band-edge emissions. The variation in PL intensity that we see in Figure 2a is, therefore, attributed to the interface quality of these samples. Our results confirm that ZnSe/Ga-terminated GaAs interface has superior quality than ZnSe/As-terminated GaAs. Moreover, the long wavelength defect peaks that are commonly observed in the PL spectra of ZnSe are minor in this study\cite{int27, int32}. This suppression may be attributed to Se vacancies present in the bulk ZnSe layers grown using ZnSe compound-source material\cite{int27}. \\
Because the realization of ZnSe-GaAs HS system requires the GaAs layers to be grown at LTs compatible to ZnSe. In this contribution, we have examined the comparative performance of ZnSe/LT-MBE GaAs (sample C in Table I), in which the GaAs buffer was grown at temperatures compatible to ZnSe growth i.e. $300^\circ$C and at a V/III ratio $\sim1$ to avoid excess As in bulk GaAs materials\cite{int28}. It can be seen that sample C indeed compromised the PL intensity greatly (see Figure 2a), further proving other epitaxy technique than regular MBE is required to develop high quality GaAs layers at LT for the realization of HS. \\
Having established that the PL intensity of ZnSe/Ga-terminated GaAs (sample A) in this study as a standard reference, we have proceeded further to develop high quality GaAs buffer layers at LTs compatible to ZnSe. In this contribution, the LT-MEE growth conditions were optimized by varying the As amount per cycle at a fixed Ga amount (1 ML). Initially, a ZnSe/GaAs sample was grown, where the LT-MEE GaAs layer was grown at $300^\circ$C and with a Ga beam flux equal to $\sim1$ ML/cycle and As$_4$ $\sim0.5$ ML/cycle\cite{int21, int22}. However, the resulting PL intensity from this sample was nearly as low as the LT-MBE GaAs buffer (sample C), although the RHEED pattern remained streaky during the growth. This might be explained by the deficient associativity of As$_4$ at LTs. In order to verify that, a set of ZnSe/GaAs samples were grown, where the LT-MEE GaAs layers were grown for various amount of As$_2$ per cycle. The normalized PL band-edge emission intensity of ZnSe obtained from these samples versus the amount of As$_2$ per cycle used, is plotted in Figure 2b. It can be seen that ZnSe PL intensity increases with the amount of As$_2$ per cycle, and reaches the optimal value when the As$_2$ dimmers supplied per cycle was precisely equal to the number of surface sites of GaAs (100) surface. From the prospective of PL performance, the optimized ZnSe/LT-MEE GaAs interface (sample D) was nearly as the same quality as the widely believed least defective ZnSe/Ga-terminated GaAs interface (sample A). Therefore, the same optimized growth conditions of LT-MEE GaAs (sample D) were used in the subsequent studies.\\

\subsubsection{\label{sec:level3}Structural and chemical investigations:}
\begin{figure*}
\includegraphics[width=0.9\textwidth]{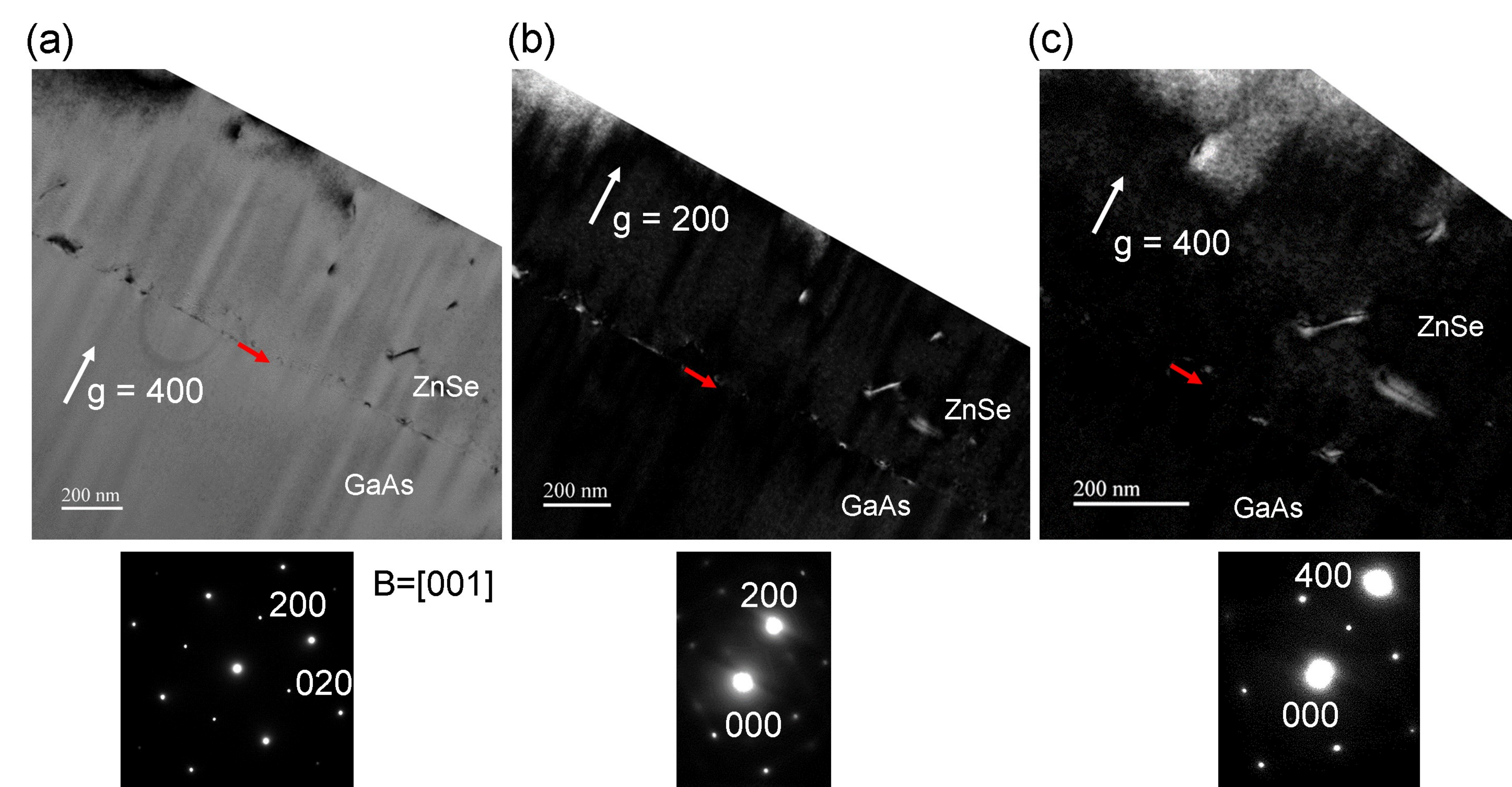}
\caption{\label{fig:epsart}Cross-sectional TEM micrographs of ZnSe/Ga-terminated GaAs hetero-structure (sample A) acquired under various operating reflections along the [001] orientation, (a) shows the microstructures of ZnSe and GaAs epi-layers obtained under two-beam bright field conditions for g = 400 reflection, (b) \& (c) shows the corresponding microstructures of ZnSe and GaAs obtained under weak-beam dark field microscopy conditions, and for g = 200 and 400 reflections.}
\end{figure*}
The bulk microstructural and interface quality of ZnSe/GaAs sample which delivered best PL response (i.e. sample A) was examined by TEM. Figure 3 shows the corresponding cross-sectional TEM micrographs along the [001] direction as indicated by the selected area diffraction (SAED) patterns. The results show that both the GaAs and ZnSe bulk epitaxial layers of sample A exhibit superior microstructural quality (see Figure 3a). The nature of ZnSe/Ga-terminated GaAs interface has been examined by weak beam dark field microscopy for the operating reflections g = 200, and g = 400 in Figures 3b and 3c. It should be noted that the weak beam dark field images do not show any complementary contrast (bright and dark) along the interface for g = 200 and 400 reflections (indicated by red arrows in Figure 3b and 3c), suggesting that the ZnSe/GaAs interface of sample A contains no signature of Ga-Se (i.e. Ga$_2$Se$_3$) compound transition layer formation at the interface.\\
With the aim of understanding further the interface structure of sample A, high-resolution TEM investigations were carried out at the ZnSe/Ga-terminated GaAs interface, however, along the [011] orientation (see Figure 4a). It can be seen that the interface is highly coherent without any misfit dislocations. The chemical composition of these layers across the interface was examined by scanning TEM coupled with the energy dispersive spectroscopy (EDS) in Figure 4b. It should be noted that the composition of ZnSe and GaAs atomic constituents varies abruptly at the interface within a thickness range of $\sim6$ nm. The Fast Fourier Transform (FFT) obtained from this ultra-thin 6 nm interface layer is provided in Figure 4a, along with the FFTs of the corresponding bulk GaAs and ZnSe layers. It can be seen that all the three FFTs are similar, and especially the FFT obtained from the interface region is not different from the FFTs obtained for ZnSe and GaAs epitaxial layers. This result suggests that the 6 nm interface layer is similar in structure to bulk ZnSe, and GaAs layers, however, chemically it is expected to contain a mixture of GaAs and ZnSe atomic constituents. The corresponding extracted lattice parameters are 0.5769 nm, 0.5702 nm, and 0.5741 nm respectively for the bulk ZnSe, interface layer, and bulk GaAs layer. \\
The TEM results in this work, therefore, suggest that the ZnSe/Ga-terminated GaAs interface (sample A) has a transition layer which is entirely different in nature from the commonly observed Ga$_2$Se$_3$ transition layer. The EDS and high-resolution TEM analyses suggest that the interface layer has a structure similar to ZnSe and GaAs, and contains all the four atomic constituents with variable concentration. Importantly, the Se counts in the EDS profile of bulk ZnSe is found to lower than Zn, suggesting that the ZnSe layers grown using the compound-source in these studies are Se-deficient\cite{int27}.\\
\begin{figure*}
\includegraphics[width=0.9\textwidth]{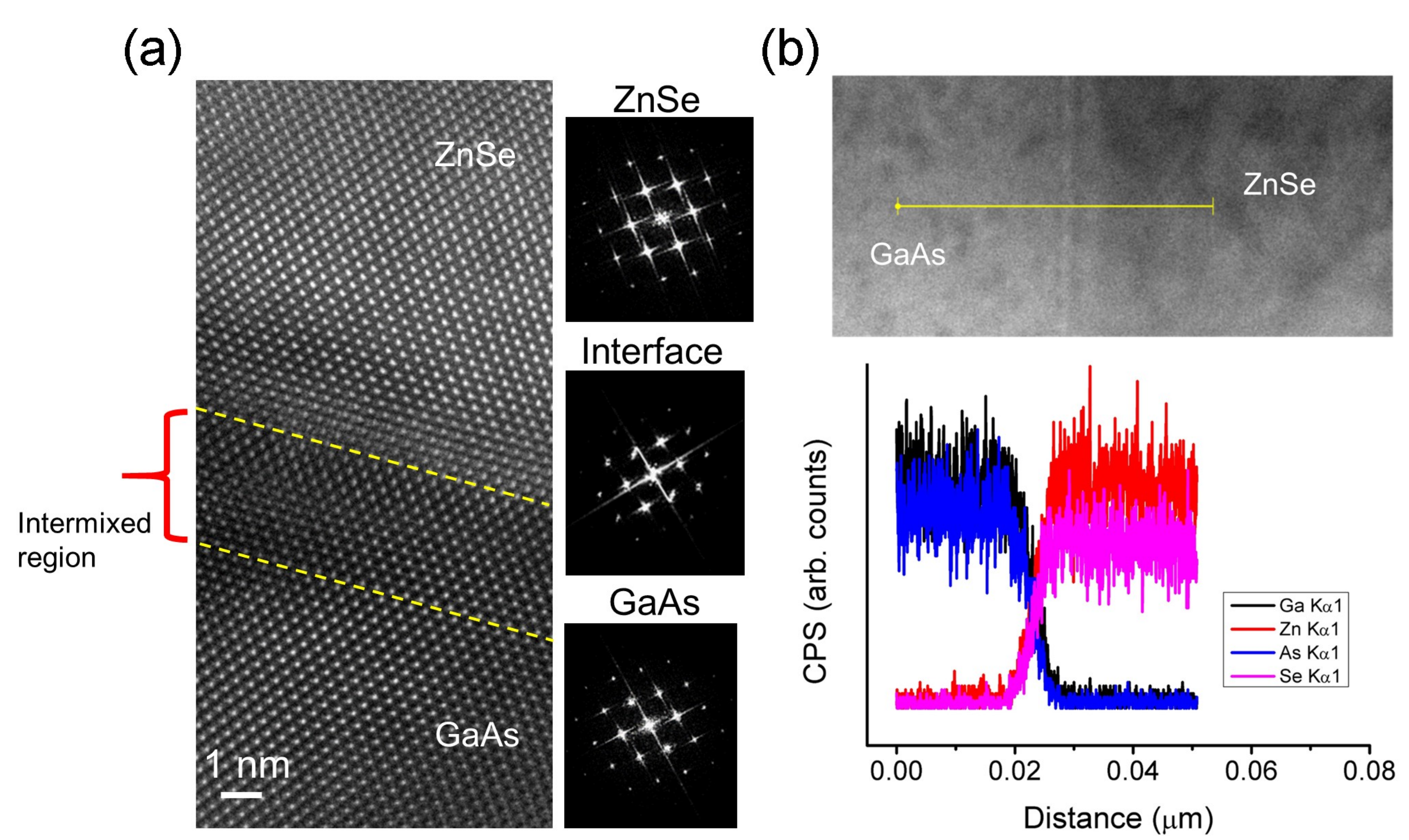}
\caption{\label{fig:epsart}(a) Cross-sectional high-resolution TEM micrograph of ZnSe/Ga-terminated GaAs hetero-structure (sample A) along [011] orientation.  The corresponding FFTs have been acquired at different regions of the image, (b) shows the scanning TEM image of the same hetero-structure along the with the EDS line scan profiles across the interface.}
\end{figure*}

\subsection{\label{sec:level2}GaAs/ZnSe Interface:}
\subsubsection{\label{sec:level3}Photoluminescence investigations:}
Having established that the LT-MEE technique can produce ZnSe/GaAs interfaces (sample D) which are comparable to the standard MBE based ZnSe/Ga-terminated GaAs (sample A), we turned our attention to investigate the LT-MEE GaAs/ZnSe interfaces. The GaAs/ZnSe interface has been identified as the bottleneck in the development of ZnSe-GaAs HS. A systematic investigation of GaAs/ZnSe interface, therefore, is essential for the development of HS. \\
In this study, we carefully controlled the initial growth layer of LT-MEE GaAs on $c(2\times2)$ Zn-rich ZnSe surface by the methods mentioned in Section II. It is observed that both the Ga-initialized GaAs/ZnSe and As-initialized GaAs/ZnSe samples have a decent ZnSe PL intensity (samples E and F in Table I). Nevertheless, it was difficult to examine the absolute value of the intensity quantitatively since the top $\sim100$ nm GaAs layers on ZnSe absorb the incident laser and ZnSe emission signals. Moreover, the top GaAs layers were too thin to produce significant PL intensity. Figure 5a shows the normalized ZnSe PL band-edge emission peak intensity as a function of incident laser power obtained for the samples E and F. The plot is a good qualitative indicator of defect density, since the defect states trap the excited carriers and influence the rate of radiative recombination. At lower incident laser powers, most of the laser energy will be consumed to fill the available defect states, provided the density of defect states is relatively fixed per sample. Consequently, the plot exhibits a strong nonlinearity if the defect states play a dominant role. From Figure 5, It can be seen that the As-initialized GaAs/ZnSe (sample F) exhibits a strong nonlinear PL response, whereas, the Ga-initialized GaAs/ZnSe (sample E) exhibits a more linear PL response.  Moreover, the absorption coefficients of ZnSe and GaAs are high for 405 nm laser, which has much higher energy than the band gaps, the corresponding defect states should be near the surface, especially the GaAs/ZnSe interface. The result in this study indicates that the As-initialized GaAs/ZnSe interface (sample F) is more defective. Besides, the RHEED pattern shows some transitional behavior from spotty to streaky for both the samples during growth (Figure 5b), suggesting that both the interfaces were not ideally abrupt and flat, and a 3D to 2D growth mode transition might have occurred during the interface formation.\\
\begin{figure*}
\includegraphics[width=0.9\textwidth]{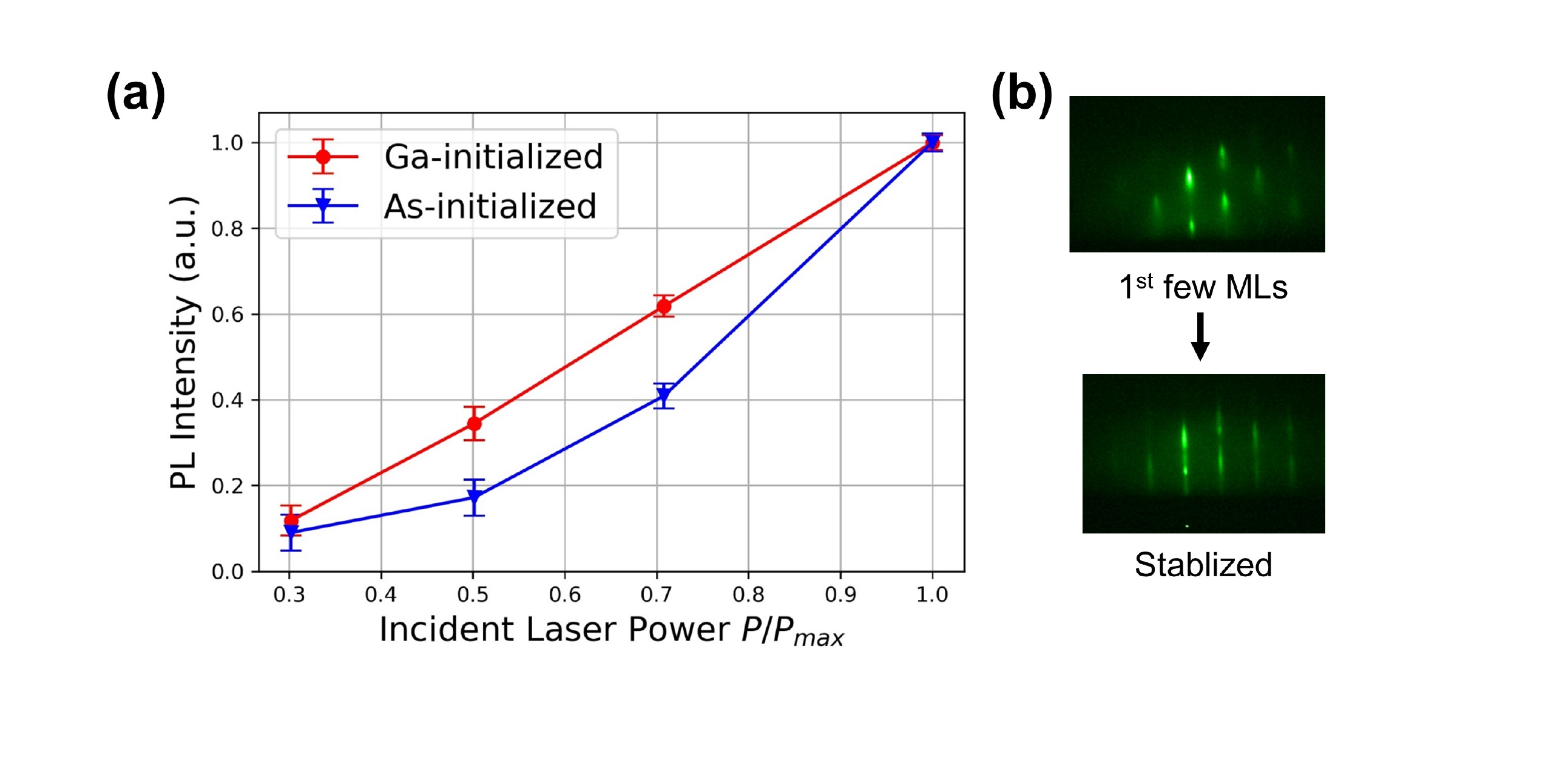}
\caption{\label{fig:epsart}Figure 5. (a) RT PL peak responses intensity versus incident laser power from ZnSe layers under different initialized MEE GaAs conditions (sample E and F) and (b) RHEED pattern during MEE GaAs growth on ZnSe}
\end{figure*}

\subsubsection{\label{sec:level3}Structural investigations:}
The structural quality of Ga and As-initialized LT-MEE GaAs/ZnSe interfaces (samples E and F in Table I) were investigated by TEM. Figure 6a and 6b shows the corresponding cross-sectional TEM bright field micrographs along the [012] orientation of GaAs/ZnSe interface, and for the operating reflection g = 400. The SAED patterns were acquired from the GaAs/ZnSe interface regions. The following observations were made from these two micrographs: the interfaces between LT-MEE GaAs and ZnSe layers are quite rough, both MEE GaAs layers are highly defective, and a defective transition region is present beneath the As-initialized GaAs layer (sample F) as indicated by the yellow colored dashed lines in Figure 6b. \\
\begin{figure*}
\includegraphics[width=0.9\textwidth]{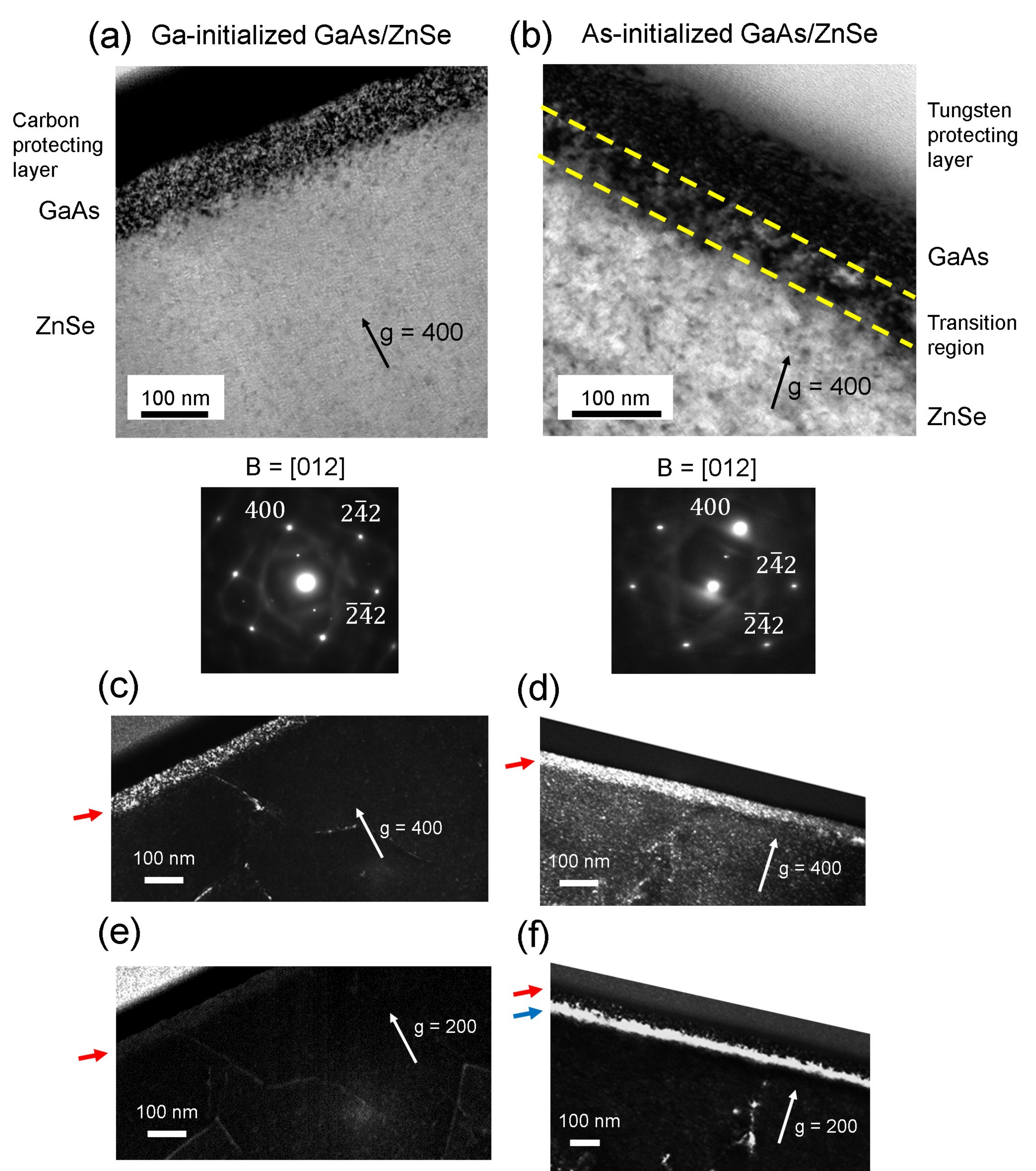}
\caption{\label{fig:epsart}Cross-sectional TEM micrographs of the Ga-initialized GaAs/ZnSe and As-initialized GaAs/ZnSe hetero-structures (samples E and F in Table I) along the [012] orientation. (a) \& (b) shows the corresponding microstructures acquired under two-beam bright field diffracting conditions and for the g = 400 reflection. (c), (d), and (e), (f) shows the images of the same hetero-structures under weak beam dark filed microscopy conditions for g = 400 and 200 operating reflections.  The red and blue colored arrows indicate the position of top GaAs layers and the position of GaAs/ZnSe interface.}
\end{figure*}
The interface quality of LT-MEE GaAs/ZnSe layers were investigated by weak beam dark field microscopy in Figures 6c to 6f. It can be seen that the top thin Ga and As-initialized GaAs layers exhibit a bright contrast for g = 400 operating reflection, and dark contrast for g = 200 reflection (indicated by red arrows in Figures 6c to 6f). It should be further noted that the As-initialized GaAs/ZnSe interface exhibits a continuous bright contrast for g = 200 reflection as indicated by blue arrow in Figure 6f, whereas the bright contrast is absent for Ga-initialized GaAs/ZnSe interface for the same g = 200 reflection (Figure 6e). Importantly, the bright contrast is absent for g = 400 reflection in both the MEE GaAs/ZnSe interface layers (Figures 6c and 6d). These observations imply that the Ga and As-initialized GaAs/ZnSe interfaces (samples E and F) are different in nature. It appears that the non-linear PL behavior of the As-initialized GaAs/ZnSe sample seen in Figure 5a, could be attributed to the defective transition region present along the interface.\\

\section{Discussion:}
\subsection{ZnSe/GaAs interface:}
Earlier CV and PS investigations revealed that the ZnSe/Ga-terminated GaAs interface has the lowest interface state density\cite{int11,int13,int14}, which was lower than the usual ZnSe/As-terminated GaAs interface obtained under As-rich MBE growth conditions. The results in the current study show that the MBE grown ZnSe/Ga-terminated GaAs hetero-structure (sample A) has the best optical as well as structural quality when compared to sample B. Many investigators studied the nature of the transition layer which forms at the interface of ZnSe/Ga-terminated GaAs\cite{int9,int29,int30}. It was reported that the transition layer has a structure similar to zinc-blende Ga$_2$Se$_3$ and exhibits a bright and dark contrast at the interface of ZnSe/GaAs for the operating reflections g = 200 and 400 in the weak beam dark field TEM\cite{int9}. In a later study, Dai et al.\cite{int31}, intentionally grown a thin Ga$_2$Se$_3$ compound layer on Ga-terminated GaAs surface and observed vacancy ordering in Ga$_2$Se$_3$ layer. The TEM results in the current study revealed no signature of Ga$_2$Se$_3$ at the ZnSe/Ga-terminated GaAs interface. The absence of complementary contrasts along the interface for g = 200 and 400 reflections in dark field TEM images confirm the same (Figures 3b \& 3c). Furthermore, the FFT analysis doesn’t reveal any spots corresponding to vacancy ordering at the interface (Figure 4). Importantly, the extracted lattice parameter of the interface (i.e. 0.5702 nm), which is higher than the lattice parameters reported for Ga$_2$Se$_3$ in literature: 0.529 nm, 0.538 nm\cite{int9, int29}, suggesting that the interface in this work has a different nature. The extracted lattice parameter of the interface is (0.5702 nm) is comparable to the lattice parameters of bulk ZnSe (0.5769 nm) and GaAs (0.5749 nm) respectively. Chemically, this interface is found to have a thickness of $\sim6$ nm with the presence of ZnSe and GaAs atomic constituents, suggesting that the interface may be an intermixture of Zn, Ga, As, and Se atoms situated in a zinc-blende structure with variable composition over a range of 6 nm in thickness. No complementary contrasts are observed for this intermixed layer for g = 200 and 400 reflections in the TEM dark field images due to the close values of atomic scattering factors associated with Zn, Ga, As, and Se atoms. It was suggested that the intermixing at ZnSe/GaAs interface could occur due to the inter-diffusion of Zn and Ga atoms during the post-growth annealing of ZnSe/GaAs layers\cite{int32}. In relation to this observation, the ZnSe growth in the current study was performed at $300^\circ$C for a relatively longer time on a Ga-rich GaAs surface. We speculate that the observed 6 nm of intermixed region in this study, therefore, may be attributed to the inter-diffusion of atomic species across the interface during ZnSe growth. \\
Moreover, the superior quality of ZnSe/Ga-terminated GaAs hetero-structure (sample A) can be rationalized from the observations proposed by Farrell et al.\cite{int20}, where it was suggested that a ZnSe/GaAs interface consists of 50\% Ga-Se and 50\% Zn-As bonds maintains the charge neutrality, thus, provide high quality ZnSe layers in 2D growth mode. The EDS observations in this study revealed the intermixing nature of ZnSe/GaAs interface (Figure 4b), however, it doesn’t provide the actual stoichiometry of the interface. On the other hand, electron counting model suggested that the bonding between ZnSe and pure Ga-terminated GaAs surface cannot directly satisfy charge neutrality without the formation of transition layers, like Ga$_2$Se$_3$\cite{int5, int20}. Since there is no trace of Ga$_2$Se$_3$ formation in this study, inter-diffusion could be an alternative way to satisfy charge neutrality. It appears that the inter-diffusion of atomic species across the ZnSe/GaAs interface resulted in a stable configuration which could yield high quality ZnSe given the growth conditions used in this study. \\
The formation of Ga$_2$Se$_3$ layer at the conventional ZnSe/GaAs interface in earlier studies may be attributed to the exposure of GaAs surface to Se beam flux prior to the growth of ZnSe\cite{int11, int14}. It was also suggested that the formation of Ga$_2$Se$_3$ can be prevented by exposing GaAs surface to Zn beam flux prior to ZnSe growth\cite{int11, int14}.  It should be noted that in earlier studies the ZnSe/GaAs hetero-structure growth was performed in two separate MBE chambers using elemental source materials as precursors for GaAs and ZnSe\cite{int13, int14}.  In these studies, prior to the growth of ZnSe, the as-grown GaAs samples were transferred to II-VI chamber and were annealed under the ambient of Se to obtain Ga-rich GaAs surface\cite{int11, int13, int14}. The reaction between Se and Ga-rich GaAs surface could result in the formation of Ga$_2$Se$_3$ compound layer on the GaAs surface\cite{int9, int13, int14}. In contrast to these earlier studies, in the current work both the GaAs and ZnSe depositions were conducted in a single MBE chamber using elemental sources as precursors for GaAs and a compound source material as a precursor for ZnSe. As a consequence, there are less chances for Se species to react with the Ga-rich GaAs surface during the annealing process and to form Ga$_2$Se$_3$. The absence of Ga$_2$Se$_3$ at the interface of ZnSe/GaAs (sample A) in our study may be attributed to the use of compound-source material for the growth of ZnSe on a Ga-rich GaAs surface in a single MBE chamber.\\
Previous studies showed that the ZnSe layers grown using compound-source material contains high concentration of impurities and Se vacancy defects\cite{int27}.  The EDS profile of bulk ZnSe shown in Figure 4b is consistent with these reports. It can be seen that the EDS profile corresponding to Se is consistently lower than the EDS profile corresponding to Zn (Figure 4b), suggesting that the ZnSe layers in this study are Se deficient. Nevertheless, the Se vacancies in ZnSe could enhance the band-edge emission and suppress the long-wavelength peaks\cite{int27, int32}. In addition, the use of compound-source material allows us to better maintain the GaAs surface chemistry by reducing the background Zn and/or Se species especially in the case of single chamber MBE systems\cite{int36}. Therefore, we believe that compound-source material could be a better choice to obtain II-VI/III-V hetero-structures with optimum quality especially when single growth chambers are used for the growth of these materials. . \\

\subsection{GaAs/ZnSe interface:}
In contrast to the well-established ZnSe/GaAs interface, the GaAs/ZnSe interface was less investigated\cite{int4, int20}, due to the difficulties involved in the deposition of this hetero-structure. For instance, ZnSe decomposes below the conventional growth temperatures used for MBE GaAs\cite{int2, int20}. In order to realize GaAs/ZnSe interface, the GaAs layers, therefore, have to be grown at temperatures as low as $300^\circ$C i.e. where the ZnSe deposition usually occurs. However, the LT-MBE GaAs layers obtained under conditions favorable to ZnSe, exhibit poor crystalline quality, and poor optical performance due to incorporation of excess As in the bulk of GaAs. The PL response obtained from the ZnSe/LT-MBE GaAs (Figure 2a, sample C in Table I) in the current study confirms the same. Moreover, the chemical valance mismatch at the GaAs/ZnSe interface adds further difficulty\cite{int4, int5, int6}. The alternative option is to use a different growth method such as MEE, which is known in producing high quality GaAs epitaxial layers at low growth temperatures\cite{int21, int22}. In MEE growth, the Ga and As precursors are supplied alternatively to enhance the mobility of Ga adatoms at such low growth temperatures, and to achieve high crystalline quality\cite{int21, int22}.  The other advantage with MEE is its ability to control the interface configuration by different initial modes. As a validation, we have employed this technique to initially produce ZnSe/LT-MEE GaAs hetero-structures (sample D) with optical repose comparable to the conventional ZnSe/Ga-terminated GaAs (sample A) and we were successful. Having established that, we have subsequently grown $\sim100$ nm thick LT-MEE GaAs layers on $c(2\times2)$ Zn-rich ZnSe surfaces (samples E and F in Table I) under different initializations of GaAs. However, the LT-MEE GaAs/ZnSe interfaces obtained were quite rough (see Figure 6). The evolution of surface roughness was monitored from the transition of RHEED patterns. During the growth of first few monolayers of GaAs for both the interfaces, the RHEED exhibited spotty pattern (Figure 5b), which can be correlated to the 3D growth mode of GaAs and hence the associated GaAs/ZnSe interface roughness. Previous studies showed that the deposition of GaAs at low temperatures using MBE on $c(2\times2)$ Zn-rich ZnSe surfaces resulted in GaAs layers with 3D growth morphology up to a thickness of 200 nm, then subsequently transformed to a 2D growth mode\cite{int20}. The rough morphology in these layers was correlated with the presence of antiphase domains\cite{int20}. However, Se pre-deposition followed by a low temperature deposition of one-half ML of Ga prior to the deposition of GaAs at $300^\circ$C, resulted in a decrease in 3D to 2D transition thickness to 20-25 nm of GaAs\cite{int20}. Though our GaAs layers were grown using MEE at $300^\circ$C, they also exhibited 3D growth mode for the first few nano meters followed by a transition to 2D mode. We didn’t measure the actual thickness at which this transition occurs, however, we have seen this transition within a minute from the time at which GaAs growth begins, which can be correlated to an approximate thickness less than 10 nm. \\
Although structurally the interfaces are quite rough, both the samples E and F still showed decent ZnSe PL response, but exhibiting different dependency on the incident laser powers (Figure 5). Our results show that the As-initialized GaAs/ZnSe sample exhibits a strong non-linearity in PL response (Figure 5a), suggesting the more defective nature of this interface. Moreover, a defective transition region with complementary contrast (bright and dark) was observed along the As-initialized GaAs/ZnSe interface for g = 200 and 400 reflections (Figure 6d and 6f). The TEM analysis in this study indicates that the transition region should have a structure factor higher than ZnSe and GaAs for g = 400 reflection, and lower than ZnSe and GaAs for g = 200 reflection. As a consequence, in dark field mode this region appears as bright for g = 200 reflection (see Figure 6f) and dark for g = 400 reflections (Figure 6d). Kuo et al.\cite{int29}, observed a similar bright contrast region along the ZnSe/GaAs interface for the conditions where the As-stabilized GaAs surface was treated with Zn. It was suggested that the transition region consists of Zn, As and vacancies as constituents and similar to Zn$_3$As$_2$ compound layer\cite{int29}. We believe that the defective transition region observed in this work along As-initialized GaAs/ZnSe interface is attributed to the formation of Zn$_3$As$_2$ compound layer.\\
The formation of Zn$_3$As$_2$ defective transition region may be explained from the process cycle used in this study (Figure 1d).  The initial As exposure at $80^\circ$C to $c(2\times2)$ Zn-rich ZnSe surface coupled with the subsequent temperature ramp up and anneal processes could induce the reaction between As-initialized GaAs and Zn-rich ZnSe surface\cite{int32}, resulting in the formation of a transition region similar to Zn$_3$As$_2$ with structure factor higher (lower) than ZnSe and GaAs for 200 (400) operating reflections (Figure 6d and 6f). A similar transition region was not observed along the Ga-initialized GaAs/ZnSe interface (Figure 6e). However, the Ga-initialized GaAs/ZnSe interface (sample E) obtained in this study was not abrupt as well. The interface roughness associated with the Ga-initialized GaAs/ZnSe interface may be explained from the reaction between the initial Ga and ZnSe surface. Previous studies show that group III (Ga, In) liquid droplets could be formed during the growth of III-V materials using MBE, and they could alter the interface properties\cite{int33, int34}. Chen et al. reported that during the growth of InAs on GaP, the initial In adlayer could form liquid droplets and dissolve GaP surface, resulting a rough interface between InAs/GaP\cite{int33}. It was also reported that Ga droplets could dissolve ZnSe during the growth of GaAs on ZnSe\cite{int34, int35}. It appears that a similar reaction might have occurred for sample E, during the Ga-initialization on Zn-rich ZnSe at $300^\circ$C, and it resulted in a rough interface between GaAs/ZnSe. 	\\
We believe that the findings in the current study may provide a better understanding of the nature of interfaces for the realization of ZnSe-GaAs HS, using the combination of MBE and MEE as growth techniques. However, further optimization of growth conditions may require to practically realize GaAs/ZnSe/GaAs interfaces with tolerable defect density. \\

\section{Conclusions:}
Systematic investigations on both the ZnSe/GaAs and GaAs/ZnSe interfaces for the realization of ZnSe-GaAs HS were performed using PL and TEM. For ZnSe/GaAs interface, the ZnSe/Ga-terminated GaAs is found to deliver excellent PL performance. Detailed TEM investigations of this interface, excluded the possibility of Ga$_2$Se$_3$ formation at the interface, instead an intermixed interface consisting of ZnSe and GaAs atomic constituents with variable composition was confirmed.  An optimized low temperature MEE procedure using As$_2$ was used to develop the GaAs layers with adequate GaAs/ZnSe interface quality, which could resolve the previous challenges involved in the deposition of GaAs at low temperatures, while maintaining the adequate interface quality. Both the Ga and As-initialized LT-MEE GaAs/ZnSe interfaces were investigated, where the As-initialized GaAs/ZnSe interface is found to be more defective and associated with a transition region corresponding to Zn$_3$As$_2$. The observed properties of ZnSe/GaAs and GaAs/ZnSe interfaces were discussed in detail in relation to the growth conditions used in this study. We believe that the findings in this study are important to understand the nature of these interfaces for the realization of high quality ZnSe-GaAs HS. \\

\begin{acknowledgments}
The authors sincerely thank Dr. Zhaoquan Zeng for valuable discussions and assistance on MBE growth.
\end{acknowledgments}

\textbf{AVAILABILITY OF DATA}\\
The raw data that support the findings of this study are available from the corresponding author upon reasonable request.\\

\nocite{*}
\bibliography{aipsamp}
\end{document}